\documentclass[draft]{aa}
\newcommand{\etal}{{\it et al.}}

\newcommand{\ltappeq}{\mathrel{\hbox{\rlap{\hbox{\lower4pt\hbox{$\sim$}}}\hbox{$<$}}}}
\newcommand{\aanda}{{\sl A\&A}}

\begin{document}

\title{On the spectroastrometric separation of binary point-source fluxes} 

\author{John M. Porter$^1$, Ren\'{e} D. Oudmaijer$^2$ \& Debbie Baines$^{2,3}$}

\offprints{jmp}
\mail{jmp@astro.livjm.ac.uk}

\institute{$^1$Astrophysics Research Institute, Liverpool John Moores University, 
Twelve Quays House, Egerton Wharf, Birkenhead, CH41 1LD, United Kingdom\\
$^2$School of Physics and Astronomy, University of Leeds, Leeds, LS2
9JT, United Kingdom\\
$^3$Astrophysics, University of Oxford, Denys Wilkinson Laboratory,
Keble Road, Oxford, OX1 3RH, United Kingdom} 

\date{Received 14 November, 2003 / Accepted 11 August, 2004}

\titlerunning{Spectroastrometry of binary point sources}
\authorrunning{John M. Porter, Ren\'{e} D. Oudmaijer \& Debbie Baines}

\abstract{
Spectroastrometry is a technique which has the potential to resolve
flux distributions on scales of milliarcseconds.
In this study, we examine the application of spectroastrometry to
binary point sources which are spatially unresolved due to the
observational point spread function convolution.
The technique uses measurements with sub-pixel accuracy of 
the position centroid of high signal-to-noise long-slit spectrum
observations.  With the objects in the binary
contributing fractionally more or less at different wavelengths
(particularly across spectral lines), the
variation of the position centroid with wavelength provides some
information on the spatial distribution of the flux.
We examine the width of the flux distribution in the spatial
direction, and present its relation to the ratio of the fluxes of the
two components of the binary. 
Measurement of three observables (total flux, position centroid and
flux distribution width) at each wavelength
allows a unique separation of the total flux into its component
parts even though the angular separation of the binary is smaller than
the observations' point-spread function.
This is because we have three relevant observables for
three unknowns (the two fluxes, and the angular separation of the
binary), which therefore generates a closed problem.
This is a wholly different technique than conventional deconvolution methods,
which produce information on angular sizes of the sampling
scale. Spectroastrometry can produce information on smaller
scales than conventional deconvolution, and is successful in
separating fluxes in a binary object with a separation of less than
one pixel.
We present an analysis of the errors involved in making binary
object spectroastrometric measurements and the separation method, and
highlight necessary observing methodology.
\keywords{
methods: data analysis -- techniques: high angular resolution,
spectroscopic -- instrumentation: spectrographs -- stars: binary: general 
}
}
\maketitle

%-----------------------------------------------------------------%
\section{Introduction}

A telescope is limited by its resolution: traditionally, the smallest
resolvable feature will typically be the size of the point-spread
function (PSF). There are several contributions to the PSF including the
Earth's atmosphere (via seeing) for ground based telescopes, and the
telescope itself (via its optics). Conventionally, if the physical
size of the flux distribution is smaller than the PSF, then
all the information on that scale is blended and hence cannot be retrieved.
For the special case of a binary object, there have been several
techniques which have been proposed to extract the
individual binary star spectra from the composite spectrum, involving 
single value decomposition  (Simon \& Sturm 1994), Fourier methods
(e.g. Hadrava 1995), or Doppler tomography (Bagnuolo \etal, 1992).

The technique of spectroastrometry  
allows the observer to gain some information on the distribution of
flux on a spatial scale smaller than the PSF. 
It was originally discussed in the
1980s by Beckers, (1982), and Christy \etal, (1983) with more recent
studies by Bailey (1998a, 1998b), Garcia
\etal, (1999), Takami \etal, (2001), Takami, Bailey \& Chrysostomou, (2003).
Given that the technique has the potential to provide information
about the flux distribution on milliarcsecond scales, it is surprising that
it has not been more widely exploited.

Conceptually, spectroastrometry is easy to grasp: it involves
taking a long slit spectrum, and relies on the observer being able
to determine the centroid of a flux distribution to a
fraction of a pixel.
The exact spatial centre of a flux distribution may vary with wavelength
if the components making up that distribution contribute differing
fractions of the total flux at a given wavelength. 
For example, 
a pair of objects in a binary system contribute different amounts of
flux at different wavelengths. Where one object has an emission
line, the position centroid of the flux will
move toward that object across that line before returning to
the continuum position. For an absorption line, the opposite occurs.
If the centroid position can be located
accurately enough, then the relative contribution of the two objects may
be calculated.
Any astronomical source with a flux distribution which is asymmetric
with respect to wavelength may be subject to the technique, such as
binary stars (e.g. Bailey 1998a, 1998b), or outflows and discs
(e.g. Takami \etal, 2003).

More familiar techniques of deconvolution of long slit spectra have
been presented by Courbin \etal, (2000).
Courbin \etal\ produce final results which
are sampled at the Nyquist frequency of the observation 
(determined by the CCD chip pixels, see the discussion in \S2 of
Magain \etal, 1998). Therefore, these techniques provide a method of
producing a {\em higher} resolution than the observations, although
they cannot provide any information on scales smaller than the Nyquist
scale. 

Lucy \& Walsh (2003, and references therein) produce an iterative
technique, which is applied to the 
extraction of stellar spectra from a (crowded) convolved image. 
This is able to separate fluxes from objects with overlapping PSF
by nominating {\it a priori} the objects which are point sources.
This procedure performs well in crowded fields, but is not sensitive
to objects so close that their separation is smaller than the PSF.

Here, we concentrate on a method of the separation of a
``discrete'' flux distribution -- a binary object -- into its
components. Our aim is to produce a method which is able to extract
the two fluxes from a binary object which is separated by less than a pixel,
(an achievement impossible with previously published methods)
whilst remaining competitive for larger separations when compared to
more traditional deconvolution techniques.

The current observational technique of spectroastrometry (photocentre
displacement) is supplemented by an investigation into the width of
the flux distribution and this provides the {\it raison d'\^{e}tre} of
this study: the addition of the information provided by the flux
distribution's width
closes the set of observables (total flux, photocentre position, and
width of flux distribution) and unknowns (the two spectra and their separation)
and thereby enables an
unique separation of the composite spectrum into its component
parts. 
We leave the problem of continuous flux distributions (e.g. a disc) until a
later study (spectroastrometry of a continuous flux distribution will 
yield the sub-pixel scale kinematics of the source). 

In \S2 the binary object flux distribution its statistical properties
are examined, before illustrating the flux separation
technique in \S3. 
To provide a practical guide for this technique, 
a discussion of errors is provided in \S4 and optimal observing
strategies are given in \S5.
Discussion and Conclusions are given in \S6 and \S7 respectively.

%-----------------------------------------------------------------%
\section{Observational statistics of the flux distribution of a
  convolved binary object}

The positional distribution of flux from any source is labelled as
${\cal F}_\lambda (x)$, where $\lambda$ is wavelength and $x$ is the
position along the slit.
If the source is now assumed to be a binary object with a separation of
many times the radius of either object, then the positional 
distribution of the flux is approximately
\begin{equation}
{\cal F}_\lambda (x) = f_{1,\lambda} \delta_{x, x_1} + f_{2,\lambda}
\delta_{x,x_2},
\end{equation}
where $f_{1,\lambda}$ and $f_{2,\lambda}$ are the
fluxes of the two 
objects located at positions $x_1$ and $x_2$ respectively and 
$\delta_{a,b}$ is the Kronecka delta function.

We imagine that the flux is observed with a spectrometer and detected
with a CCD detector with dimensions $n \times m$ pixels.
The flux is sampled at discrete positions and wavelength: the pixel
number in 
the spatial direction is labelled with index $i$, with $i=0$
corresponding to the bottom of the CCD frame and $i = n$ corresponding to
the top of the frame. Likewise the pixels in the dispersion (wavelength)
direction are labelled with index $j$ with $0 \leq j \leq m$.
The flux in each CCD pixel is labelled $F_{i,j}$.

During an observation, the flux distribution ${\cal F}_\lambda(x)$
is first convolved with a function taking into account 
the seeing profile (for ground based telescopes), and telescope
optics. We label this intrinsic 
``seeing'' function $S_\lambda(x)$. The flux is then binned into pixels as
it is detected by the CCD. The measured distribution $F_{i,j}$ is
then  
\begin{equation}
F_{i,j} = {\rm discretize}\left[{\cal F}_\lambda(x) \otimes S_\lambda(x)\right].
\end{equation}
We label the intrinsic width of the seeing function as $\sigma_S$,
which in the case of a Gaussian function is equivalent to the standard deviation.
If the separation between the objects $| x_1 - x_2 |$ is less than
$\sigma_S$, then the two objects are formally not
resolved. Conventionally, separation of the individual spectra cannot
be achieved. 

Observations have two main sources of noise (i) photon counting
errors, characterised by Poisson statistics of the number of photons
in each pixel, and (ii) read noise generated in the process of charge
transfer in the CCD during the read out stage. The relative
contribution of these two error sources is dependent on the exposure
time and brightness of the source observed, as well as the
spectrometer itself.
First we consider the case of zero noise to develop
the fundamental principles and technique, and we return to the effects
of noise later.

The three simple observables which may be measured from the flux
distribution on the CCD are
the total flux, the mean position of the flux, and the width of the 
flux distribution. In order to make the discussion as general as
possible, henceforth we work with units of pixels.

% ----------------------- total flux -----------------------------
\subsection{Total flux}
At a given wavelength $\lambda$, corresponding to pixels ($i=1\cdots n,j$)
the flux is distributed in position 
according to eq.1 and sampled onto pixel row $i$ of the CCD.
The ``extracted'' total flux is the sum over all
pixels in the spatial direction of the flux in each pixel:
$\sum_{i=0}^{n} F_{i,j} \equiv f_{{\rm tot},j} =
f_{1,j} + f_{2,j}$
%\begin{equation}
%E_j = \sum_{i=0}^{n} F_{i,j} \equiv f_{{\rm tot},\lambda} =
%f_{1,\lambda} + f_{2,\lambda} 
%\end{equation}
In traditional spectrographic work, this total extracted flux
$f_{{\rm tot},j}$ (the spectrum) is the sole quantity 
gleaned from the observation.

% --------------------- centroid  -------------------------------
\subsection{Position centroid}

The position of a distribution is often measured by the mean of the
distribution -- the 1$^{\rm st}$ moment. However, it is often the case
that the mean is not particularly robust for non-Gaussian
distributions (especially when noise is present, see e.g. Beers \etal,
1990), and so a more
sophisticated approach is warranted.
Good choices are the
M-estimators such as Tukey's biweight, or Andrew's Sine (or wave) estimators
(Goodall 1983, Press \etal, 1986). These have a weighting function
which penalises outliers from the distribution, which in this context
translates to estimating the width over a specific pixel region. This
will then avoid problems with extended wings of seeing profiles (and
also cosmic ray hits, and noise-generated errors).

We denote the position centroid of the distribution (however it is
measured) as
$\mu_j$: note that it can be measured to fractions of a pixel hence $\mu_j$
is a real number and not an integer. 
How does the centroid $\mu_j$ vary as the ratio of the flux of the
two objects varies? This is the aspect of spectroastrometry which has
received the most attention (Bailey, 1998a, 1998b, Garcia \etal, 1999,
Takami \etal, 2001, 2003). 
For equal fluxes $f_{1,j} = f_{2,j}$
the centroid will lie exactly in the middle of flux distribution 
(at $(x_2-x_1)$/2).

The position centroid $\mu_j$ is a direct probe of the relative
contribution of the two fluxes to the total flux.
A simple relation can be written to calculate $\mu_j$ in
terms of the two fluxes:
\begin{equation}
\mu_j \equiv \frac{f_{1,j}\delta_{x,x_1}}{f_{{\rm tot},j}} + 
\frac{f_{2,j}\delta_{x,x_2}}{f_{{\rm tot},j}},
\end{equation}
(e.g. Bailey, 1998a) assuming that both point sources are in the
spectrometer slit.

A particularly useful spectral feature which is used in the deconvolution
technique below is an emission/absorption line. These push the
centroid toward the brighter object over a small wavelength range --
clear in the centroid spectra of Mira (fig.2 of Bailey 1998a) and
HK Ori (fig.9 of Baines \etal, 2004).
Any slowly varying changes in the centroid can be subtracted
out producing a centroid spectrum $\mu_j^\prime$
which is calibrated to be zero away from the line centre (in the
continuum) and may be written as:
\begin{equation}
\frac{\mu_j^\prime}{d} \equiv 
\frac{f_{2,j}}{f_{tot,j}} - 
\frac{f_{2, {\rm cont}}}{f_{tot, {\rm cont}}} = \frac{r_j}{r_j + 1} -
\frac{r_{\rm cont}}{r_{\rm cont} + 1},
\end{equation}
(see eq.3 of Takami \etal, 2003),
where $f_{2, {\rm cont}}$ and $f_{tot, {\rm cont}}$ are the continuum
fluxes of object 2 and the total flux, and $d$ is the separation of the objects
in pixels. The final equality in eq.4 is expressed in terms of the
ratio of the fluxes $r_j = f_{2,j}/f_{1,j}$ and the flux ratio in the
continuum $r_{\rm cont} = f_{2,{\rm cont}}/f_{1,{\rm cont}}$ which is
used later in \S3.

% -------------------------- FWHM -------------------------------
\subsection{Width of the flux distribution}

The width of a distribution may be measured using several methods:
the simplest characteristic width of the flux distribution is the mean
absolute deviation, and the most commonly used is the standard
deviation. However, as with the position above, a more robust
technique is required in general.

We denote 
the measured width of the distribution as $\sigma_j$. This will
approach the limiting value $\sigma_S$ of the seeing function
$S_\lambda(x)$ as either (i)  
the separation of the two objects tends to zero, or (ii) the ratio of
the less bright object flux to the brighter object flux tends to zero.
If either of the objects becomes brighter/dimmer, then the observed
width will change. The width $\sigma_j$ will
also change with the separation of the objects.
If the separation of the objects is larger than the intrinsic
$\sigma_S$ of the seeing, then the two objects will start to be
resolved, and the spectroastrometric technique is not required to
separate the objects' spectra. Hence the width $\sigma_j$ is a function
of (i) the separation of the two objects, (ii) the intrinsic distribution's
$\sigma_S$, and (iii) the flux ratio of the two objects. 

A slight broadening of the convolved profile,
i.e. $\sigma_j>\sigma_S$, is present even in the continuum, as
both stars still contribute to the flux distribution.
This is important in that the measured $\sigma_j$ of the
continuum cannot be used as an estimate for $\sigma_S$. 
A separate measurement for $\sigma_S$ will be necessary (by observing a
single star), particularly as the focus of the camera can change along
the CCD producing a varying $\sigma_S$ in the dispersion direction.
It may be possible to use the minimum of $\sigma_j$ as an estimate for
$\sigma_S$ where there is a particularly strong emission line present
from one of the objects such that $f_{1,j} \gg f_{2,j}$, although this
will still be a slight overestimate for $\sigma_S$ (see Appendix A.2
for an evaluation of the errors associated with this procedure).

\begin{figure}
\vspace{7.8cm}
\includegraphics{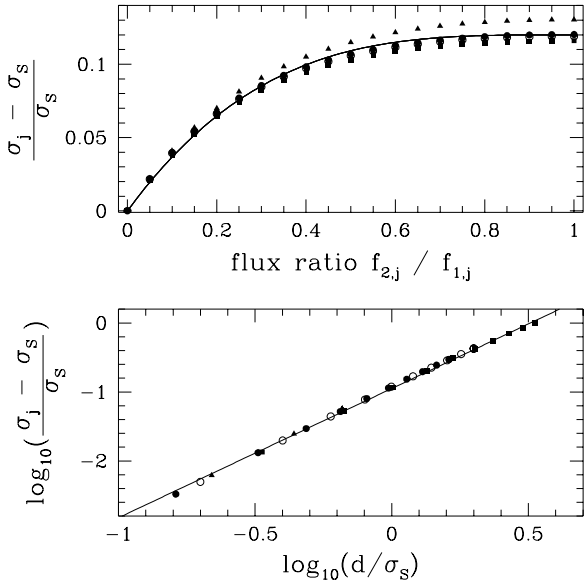}
\caption{Top panel: variation of the width of the distribution with flux
  ratio. The symbols are the calculated data for an intrinsic
  convolving function of a Gaussian (open circles, width 5.0\,pixels),
  a Gaussian with boosted wings (filled circles, width 6.2\,pixels, see text)
  a top-hat (squares, width
  3.0\,pixel), and a triangle (triangles, width 4.6\,pixels). 
  The separation between the sources is 3.0\,pixels.
  All results have been divided by $(d/\sigma_S)^b$
  ($b$ from eq.5 listed in Table 1) to show the dependency of the
  width on the flux ratio only.
  Bottom panel: variation of the width with separation $d$ between the
  sources for a flux ratio of $f_{2,j}/f_{1,j} = 0.5$. Symbols are the
  same as for the top panel. 
  The solid lines in both panels are the fit of eq.5.
}
\end{figure}

In order to understand the behaviour of the changes in width of 
the flux distribution $\sigma_j$, with flux ratio, binary object
separation and intrinsic width $\sigma_S$, we have performed an
extensive series of numerical simulations.
The flux ratio ranged from
$f_{2,j}/f_{1,j} = 0.0$--$1.0$ (stepsize 0.05); the separation ranged from $d =
0$--$3\sigma_S$ (stepsize 0.1$\sigma_S$), and we have varied the
intrinsic width $\sigma_S$ from 2--10\,pixels (similar to typical
observations).
Each realisation first convolves the binary
object flux distribution with the function $S_\lambda(x)$, and then maps
the resultant flux onto pixels.
We have also performed the calculation for different functions
$S_\lambda(x)$, including a Gaussian, a Gaussian with boosted wings
(the Gaussian function multiplied by $1 + 0.05(x/\sigma_S)^4$),
top-hat and triangular functions. In all cases we have measured the
width of the resultant distribution using Andrew's sine estimator,
although identical results are obtained with a bi-weight estimator, or
a standard-deviation width.

The results of some of these calculations are
displayed in Fig.1. We find that the behaviour of the fractional change of the width of
the distribution is almost independent of the function $S_\lambda(x)$
and may be approximated by a fitting function. 
For objects with separations less than twice the
intrinsic $\sigma_S$ (i.e. $d < 2\sigma_S$) an excellent fit to the
calculations is
\begin{equation}
\frac{\sigma_j\! -\! \sigma_S}{\sigma_S} =\! a \left(
\frac{d}{\sigma_S}\right)^{b}\! \left[ 1 - \left| 1\! -\! 
{\rm min}\left\{\! \left(\frac{f_{2,j}}{f_{1,j}}\right)\!\!,\!
\left(\frac{f_{1,j}}{f_{2,j}}\!\right)
\right\}
\right|^{c} \right].
\end{equation}
The best fitting constants $a,\ b,$ and $c$ (with a
search stepsize of 0.001) in this expression are found to 
be slightly dependent on the exact shape of the convolving function
$S_\lambda(x)$. The best fitting values are listed in Table.1.

\begin{table}
\begin{tabular}{lrrr}
$S_\lambda(x)$              &  $a$   &   $b$  &   $c$  \\ \hline
Gaussian                    & 0.116 & 1.930  & 3.468 \\
Gaussian with boosted wings & 0.118 & 1.951  & 3.468 \\
Top hat                     & 0.110 & 1.858  & 3.466 \\
Triangular                  & 0.130 & 2.012  & 3.448 \\
\end{tabular}
\caption{Best fitting parameters to eq.5, $S_\lambda(x)$ is the
  convolving function -- see text.}
\end{table}

This expression is shown in Fig.1 for different functions $S_\lambda(x)$.
Eq.5 reproduces the numerical results within a few per cent of the
calculated value for all of the convolution functions $S_\lambda(x)$
we have used (although it is slightly worse for the triangular
function than the other functions). 
The error associated with use of this fit produces
a systematic shift in the separated fluxes (especially for the
dimmer secondary object) in the practical application of the technique
(see \S3). To reduce this error, we can actually measure the
function $S_\lambda(x)$, again via a single object observation, and then
use it to calculate either more accurate values of $a$, $b$, and $c$ in
Eq.5, or a ``lookup'' table of values for use in flux separation.
Eq.5 is important in that this representation of the change in the
width contains all of the convolution information of the binary
flux distribution. 

%-----------------------------------------------------------------%
\section{Method of extraction of the individual spectra}

\begin{figure*}
\vspace*{17cm}
\includegraphics{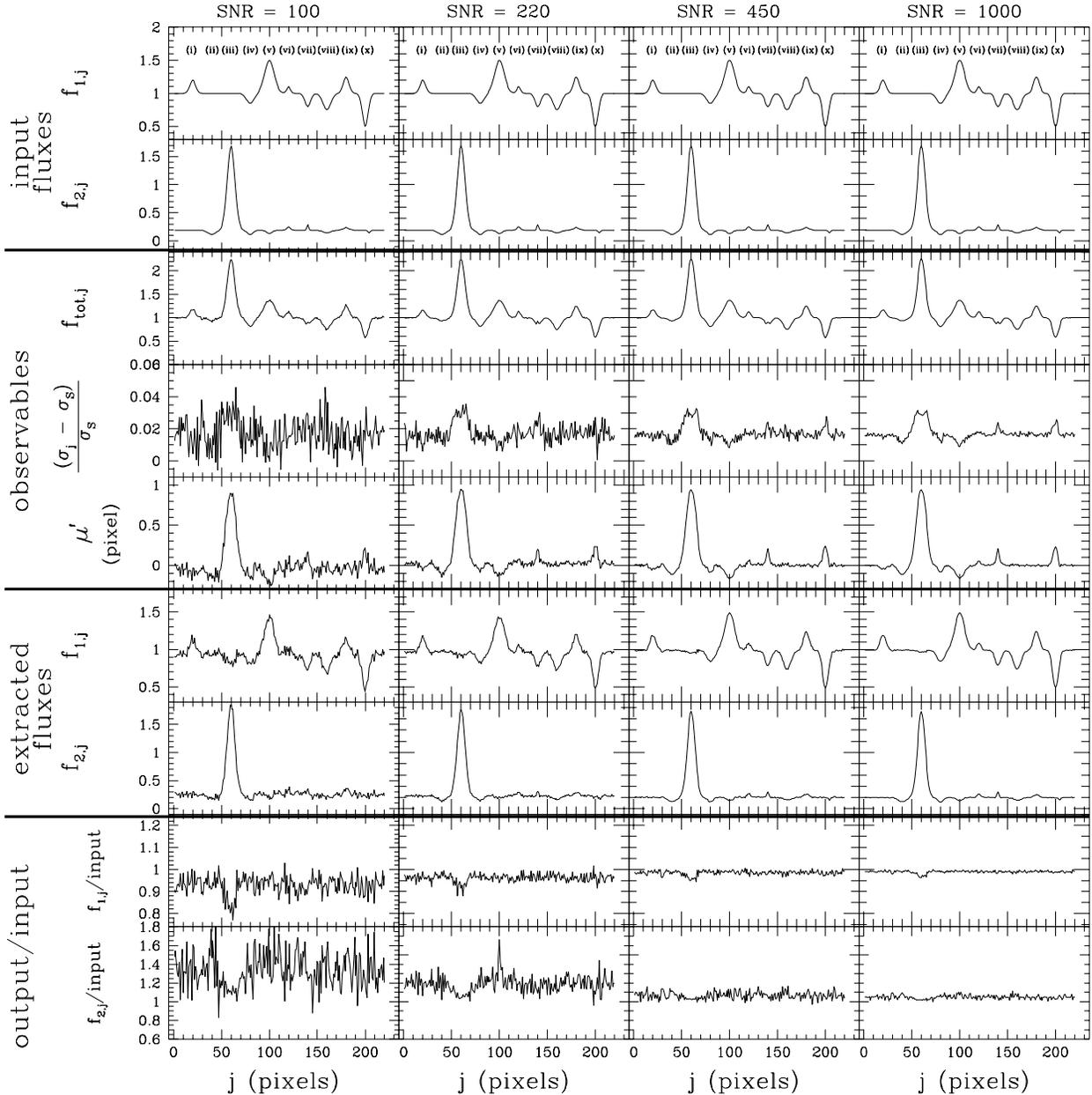}
\caption{Example A: the top two panels are the input
  fluxes of the individual objects (the primary object's flux is normalised to unity in
  the continuum).
The next three panels are the observable quantities: the total flux,
  the relative change in width of the distribution, and the change in 
position centroid.
The next two panels are the best-fit extractions of the two fluxes,
  and the bottom two panels are the ratio of the extracted flux to the
  input fluxes from the top two panels. The separation of the binary
  is $d =2.0$\,pixels, and the convolving function is Gaussian with width
  $\sigma_S = 4.0$\,pixels.}
\end{figure*}

The total flux $f_{\rm tot, j}$ is dependent on the fluxes of the two
point sources $f_{1,j}$ and $f_{2,j}$. The position displacement
$\mu_j^\prime$ is dependent on $f_{1,j}$, $f_{2,j}$ and the separation
$d$. Finally, the width of the flux distribution $\sigma_j$ is
dependent on $f_{1,j}$, $f_{2,j}$ and $d$.
Therefore measurement of (i) the
total flux $f_{\rm tot, j}$, (ii) the position displacement
$\mu_j^\prime$ and (iii) the width of the flux distributed in the
spatial direction $\sigma_j$ provides three observables for three
unknowns (i.e. the fluxes of the two objects $f_{1,j}$, $f_{2,j}$, and
their separation $d$). The set of three relations may be inverted to
uniquely separate the flux distribution into its components.

Previous attempts at spectroastrometric flux separation
have used the total flux $f_{\rm tot, j}$ and position displacement
$\mu^\prime_j$ as their only two observables.
To be successful, prior knowledge of the objects must be obtained
(e.g. Bailey 1998a knew the separation $d$ of the binary). If this is
not possible, then we have three unknowns (the two fluxes and
source separation) and only the two observables.
Hence flux separation may not be uniquely achieved as we do not
have a closed set: attempts to separate the flux using only two
observables may produce misleading results. 

We use the following method to deconvolve the spectra: first 
for a given value of the separation $d$, we calculate the 
flux ratio in the continuum $r_{\rm cont}$ with eq.5 and the observed
width $\sigma_{\rm cont}$. Then for each row in the dispersion
direction $j$ we invert eq.4 using the observed centroid $\mu^\prime_j$ and
continuum ratio $r_{\rm cont}$ to produce the flux ratio $r_j$. 
Then with $f_{\rm tot,j} = f_{1,j}+f_{2,j}$, we calculate the
individual spectra $f_{1,j}$ and $f_{2,j}$. 
Using these values, we predict the width of the distribution 
using eq.5, and compare to the observed
values using a simple $\chi^2$ calculation to evaluate the fit.
Finally we repeat this procedure with differing values of the
separation $d$ until a best fit is found.

We follow this procedure for two examples (see below). This suggested
method applies in the case that the width $\sigma_S$ of the convolving
function is known.
\footnote{We note that this procedure may be changed if the prior information is
varied: if the separation is accurately known then a similar method
could be devised which had the width $\sigma_S$ as the independent
parameter.}
The separation
$d$ is the independent parameter which is varied to minimise
$\chi^2$ for the fit.

\subsection{Example A}

In order to fully test the method, a series of emission and absorption
``lines'' of differing contrasts and widths
are imposed on the continuum of both objects (see the top two
panels in Fig.2). These provide parts of the spectrum 
which have all combinations of the primary and secondary object
brightening and dimming over the lines: from the left (in increasing
pixels) the features correspond to (i) object 1 emission line only, (ii)
object 2 absorption only, (iii) large emission line in object 2 such
that it is brighter than object 1, 
(iv) both objects absorption, (v) object 1
emission, object 2 absorption, (vi) both objects emission (vii) object 1
absorption, object 2 emission, (viii) both objects absorption,
but with the absorption minima offset, (ix) both objects emission with
identical contrast and finally (x) both objects
with absorption lines of identical contrast.
Note that the two fluxes are not meant to represent any sort of object
or binary system in particular: they are simply meant to illustrate
the differing combinations of primary and secondary flux.
The continuum flux ratio is constant at all wavelengths at 
$r_{\rm cont} = 0.2$. 

\begin{figure}
\vspace{7.8cm}
\includegraphics{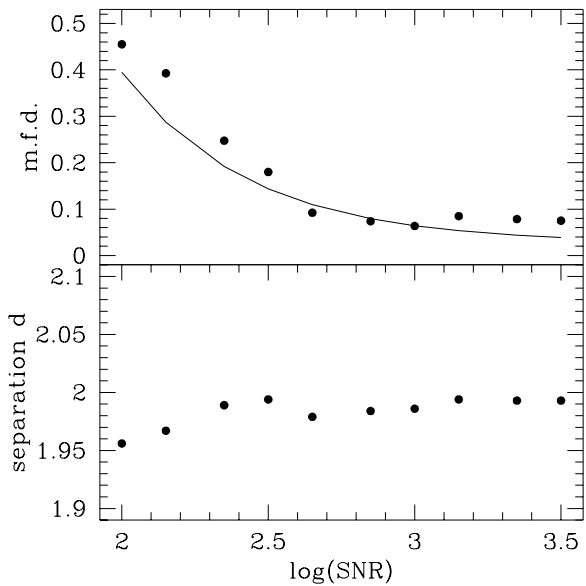}
\caption{Variation of mean fractional deviation per pixel of the
  extracted fluxes from the input fluxes (eq.6) (top panel), and
  best-fit separation (bottom panel) for Example A with increasing
  signal-to-noise ratio (SNR).
  The solid line in the top panel is the mean fractional deviation per pixel
  expected for this example, with systematic error of 1\% see \S4.
}
\end{figure}

\begin{figure}
\vspace{7.8cm}
\includegraphics{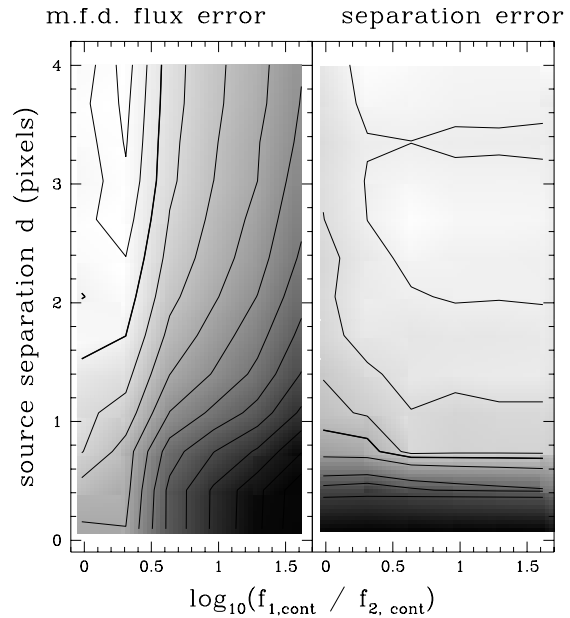}
\caption{Map of mean fractional deviation per pixel (left) and error in
  the best-fit separation (right) 
for a set of simulations using the input spectra of Example A, but
  varying the flux ratios and the object separation.
The contours are logarithmically spaced with steps of 
0.2dex. The bold contours correspond to an error of 0.1 (i.e. 10\%).
In the left panel, the highest contour corresponds to an 4.0\%
  error and is situated in the top left corner. In the right panel the
  highest contour corresponds to 2.6\% and is in the middle--top right of the
  panel.
}
\end{figure}

An ``observation'' is then made: this consists of 
convolving the input with $S_\lambda(x)$ assumed to be a
Gaussian with a width (identical to the standard deviation in this
case) of $\sigma_S = 4.0$\,pixels.
The resultant distribution is then binned into pixels in position and
dispersion.  
The observables are then calculated -- the total flux, the width, and
position offset. To derive the relative change in width, one of the
input fluxes is set to zero, and an ``observation'' taken of a single
flux component, and the width calculated.
The derived observables for a source separation of 2.0\,pixels are
displayed in the third--fifth panels of Fig.2 for a range of
signal-to-noise ratios (SNR).

One feature in the width distribution is particularly worthy of
note. When the fainter object in the continuum has a large emission
line -- feature (iii) in fig.2 -- the width of the flux distribution
$\sigma_j$ can be seen to decrease in the centre of the line,
producing a double-peaked profile. 
The distribution width $\sigma_j$ has a maximum for a flux ratio of
unity. 
How does the width $\sigma_j$ change from the line wing of feature
(iii) to line centre? In the wing, object 2 is much less bright
than object 1, 
and the flux ratio in eq.5 is less than unity. As the line emission
increases, the flux ratio increases (and hence so does the width
$\sigma_j$) until the emission from both objects is equal and the flux
ratio is 
unity. 
Here, the width of the distribution $\sigma_j$ reaches its
maximum value. 
As the line emission increases further such that object
2 is brighter than object 1, the width $\sigma_j$
{\em decreases} because $\sigma_j$ is dependent on the {\em minimum} of the
ratios $r_j$ and $1/r_j$ (this is because there must be no difference
in $\sigma_j$ whichever of the two objects is denoted object 1).

This produces a maximum in the width $\sigma_j$ which is
offset in wavelength from the maximum in emission, and hence a double
peaked profile in $\sigma_j$ is observed. The peaks in $\sigma_j$
then correspond to a flux ratio of unity.

The bottom four panels of Fig.2 show the results of the
separation method. Panels six and seven display the extracted fluxes,
and panels eight and nine show the extracted fluxes divided by the
input fluxes (if the extraction is perfect, then these panels should
be unity). It is clear from examination of these four panels that the
flux separation method achieves good results even for SNR of 100, and
that as SNR increases, the accuracy of the flux separation becomes
more accurate.
To find the best-fit, the distance between the components $d$ was
varied from 0--5\,pixels with a stepsize of 0.001\,pixels.
This procedure yielded $d=1.956,\ 1.989,\ 1.979$ and 1.986\,pixels for
SNR of 100, 220, 450, and 1000 respectively. 

To assess how well the flux separation was achieved we calculate a
mean fractional deviation per pixel of the extracted flux:
\begin{equation}
{\rm  m.f.d.} =
\frac{1}{m} \sum_{j=0}^{j=m} 
\left[\ \ 
\left| \frac{f_{1,j,{\rm ex}}}{ f_{1,j,{\rm in}}} - 1 \right| + 
\left| \frac{f_{2,j,{\rm ex}}}{ f_{2,j,{\rm in}}} - 1 \right| \ \ \right]
\end{equation}
where the subscript ``ex'' and ``in'' refer to the extracted and input
fluxes respectively.
The m.f.d. per pixel for Example A along with the best-fit
separation $d$ 
as a function of SNR are shown in Fig.3. The separation
asymptotes to the input separation of 2.0\,pixels, and the m.f.d.
per pixel decreases with increasing SNR. 
However, even for SNR$> 1000$,
the separation is not exact, and indeed, is not expected -- this is an
empirical method, and hence will contain errors (see \S4
below). The solid line on the top panel of Fig.3 is the expected m.f.d.
per pixel for this simulation as described below in \S4.

Figs.2 \& 3 illustrate the performance of the method as the SNR varies
for a constant continuum flux ratio. 
We also investigate the method for differing continuum flux ratios
$r_{\rm cont}$ and distance between the objects $d$. 
For this we have conducted a series of simulations for the same
features in the spectra (top two panels in Fig.2).
However, the continuum flux ratio $r_{\rm cont}= f_{2,{\rm cont}}/f_{1, {\rm
    cont}}$ is varied from unity to 1/50, and the separation $d$ of the
sources ranges from 0.1\,pixels to 4.0\,pixels (i.e. 0.025--1.0 of the
standard deviation of the Gaussian convolving function).
For all of these simulations, the noise level is fixed to produce
SNR$=220$ (similar to that in the simulations of Courbin \etal, 2000
for comparison of the two methods).

For each pair of $r_{\rm cont}$ and $d$ we calculate the m.f.d. per pixel
of the extracted fluxes and plot this as a contour map in the left
panel of Fig.4. The contours are spaced logarithmically in intervals
of 0.2, and the bold contour marks the m.f.d. per pixel of 0.1
(log(m.f.d.) = -1). We can clearly see that the technique performs best for large
separations $d$ and for flux ratios $r_{\rm cont}$ close to unity. We also
note that accurate extraction of the fluxes can be achieved for object
separations of less than one pixel (a feature of the technique which,
to our knowledge, is unique). 
We have also calculated the fractional error in the best-fit
separation and show the contour map of this in the right hand panel of
Fig.4. Again the contours are spaced in 0.2dex, and the bold line
corresponds to a fractional error of 0.1 (i.e. 10\%).
The main feature of this map is that the best-fit separation $d$ is
not very sensitive to the continuum flux ratio, and that accuracy of
better than 10\% should be possible for input separations of larger
than around one pixel.

\begin{figure*}
\vspace*{17cm}
\includegraphics{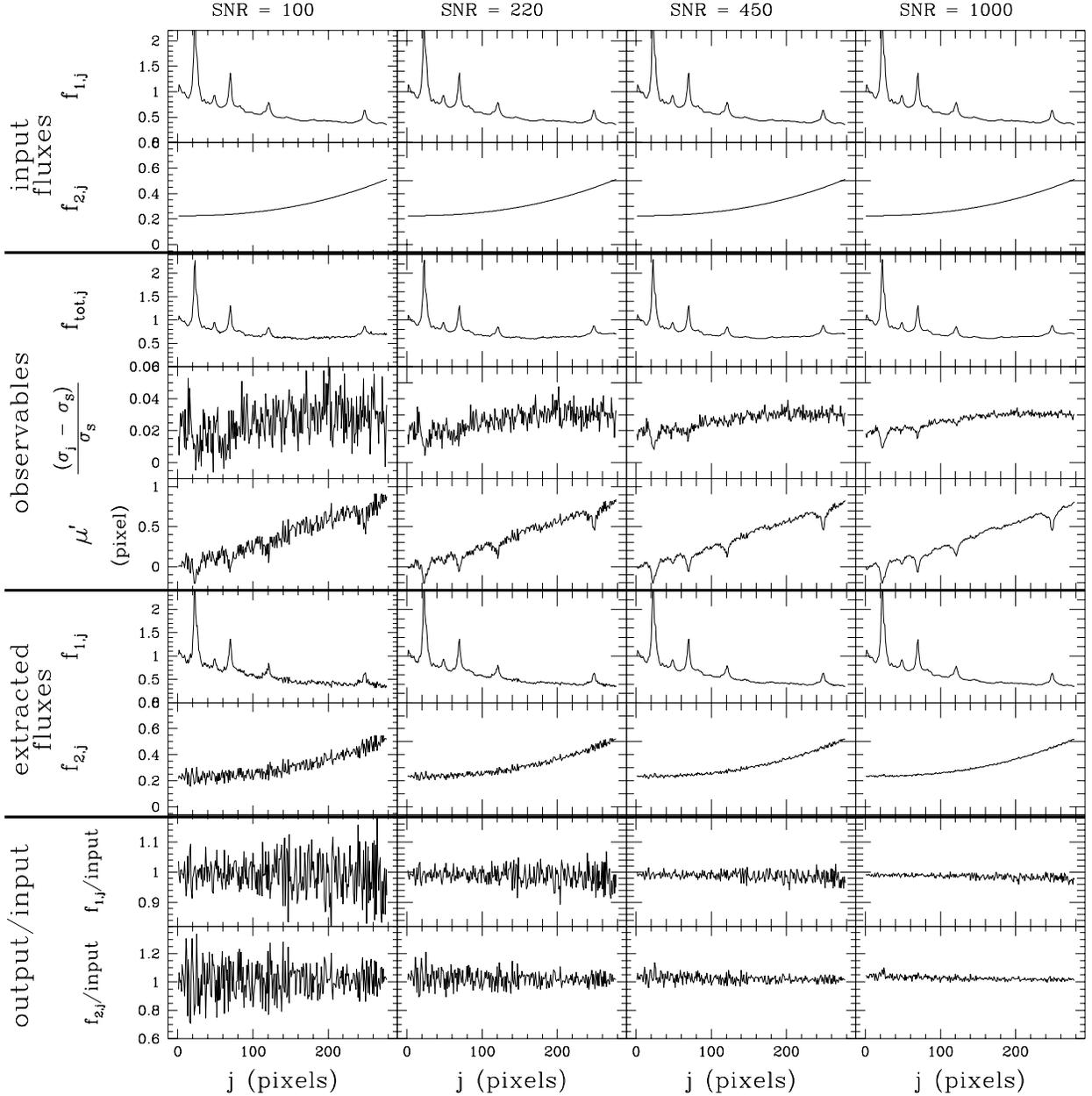}
\caption{Example B: the top two panels are the input
  fluxes of the individual objects.
The next three panels are the observable quantities: the total flux,
  the relative change in width of the distribution, and the change in 
position centroid.
The next two panels are the best-fit extractions of the two fluxes,
  and the bottom two panels are the ratio of the extracted flux to the
  input fluxes from the top two panels. The separation of the binary
  is $d =2.0$\,pixels, and the convolving function is Gaussian with width
  $\sigma_S = 4.0$\,pixels.}
\end{figure*}

\subsection{Example B}
To provide a direct comparison with previous methods, we adopt exactly
the same problem considered by Courbin \etal, (2000) for a second test.
This consists of a quasar and a star with a featureless continuum (see
Courbin \etal, 2000, \S3.1). The quasar spectrum used is the mean of all
objects taken from the 2QZ project\footnote{http://www.2dfquasar.org/}
(Croom \etal, 2004).
The two point sources are separated by
2.0\,pixels, and the convolving Gaussian to provide the seeing has a
width of 4.0\,pixels. The relative continuum brightness ranges from
$r_{\rm cont} = f_{2,{\rm cont}}/f_{1,{\rm cont}} =$0.2--1 from the
blue to the red end of the spectra.
The mean flux ratio of the two components is larger in
this case than in Example A, and therefore we 
should expect more accurate results.

Fig.5 shows the results for this test in a similar fashion as Fig.2,
and Fig.6 shows how the m.f.d. per pixel in the extracted
fluxes and the best-fit separation vary with SNR.
As can be seen, the extraction does indeed produce more accurate
results than for Example A. For the cases of SNR = 200-300 (exactly the
same as in Courbin \etal) the m.f.d. per pixel is $\approx
0.05$. We estimate that this is larger than the typical m.f.d. in
the results of Courbin \etal\ (from the inserts in their Fig.3) by a factor
of $\sim 2$.
This indicates that the spectroastrometric method of extracting fluxes 
is not quite as efficient as Courbin \etal's method for this particular
example -- unfortunately Courbin \etal\ do not discuss differing SNR
simulations. 
 
As in Example A, we examine the technique for differing values of
continuum ratios and component separations (the continuum ratio is defined
by the fluxes at pixel $j = 1$).
Fig.7 displays the contour map of m.f.d. per pixel and best-fit
separation (as Fig.4 for Example A), for variation of the
continuum ratio (unity -- 1/50) and component separation ($d =
0.1$--4.0\,pixels). We have fixed the SNR at 220.
Again, the most accurate results are obtained for larger component
separations $d$, and for larger flux ratios.

%-----------------------------------------------------------------%
\section{Random and systematic errors}

We can calculate the expected fractional deviation per pixel in
a straightforward way.
At each pixel $j$, small variations in the fluxes $f_{1,j}$ and
$f_{2,j}$ ($\delta f_{1,j}$ and $\delta f_{2,j}$ respectively)
produce a variation in the flux ratio $\delta r_j$, where 
$\delta r_j/r_j = \delta f_{2,j}/f_{2,j} - \delta f_{1,j}/f_{1,j}$.
As the sum of the two fluxes is a constant for any $j$, then the small variations
$\delta f_{1,j}$ and $\delta f_{2,j}$ are equal and opposite in sign
($\delta f_{1,j} = -\delta f_{2,j}$). Hence, $\delta r_j / r_j =
- ( 1 + 1/r_j) \delta f_{1,j}/f_{1,j}$.

The mean fractional deviation per pixel (as defined in
 eq.6) is the sum of the absolute fractional errors:
\begin{equation}
{\rm m.f.d.} = \frac{ | \delta f_{1,j} | }{f_{1,j}} + 
\frac{ |\delta f_{2,j} |}{f_{2,j}} = 
\frac{ |\delta f_{1,j} |}{f_{1,j}} \left( 1 + \frac{1}{r_j} \right) 
= \frac{ | \delta r_j |}{r_j} 
\end{equation}

In the limit that the flux ratio $r_j \ll 1$, the expected error in
$r_j$ is calculated in Appendix A.1 to be 
\begin{equation}
\left| \delta r_j \right| \approx \frac{1}{ac\ \sqrt{2}{\rm SNR}}
\left(\frac{\sigma_S}{d}\right)^b
 + \left. \delta r \right| _{\rm syst} 
\end{equation}
where $a$, $b$, and $c$ are the constants in eq.5, and 
$\left. \delta r\right|_{\rm syst}$ is the systematic error
from the fitting formula eq.5

We can use this estimate to calculate the expected m.f.d. error for
Examples A \& B. Inserting a systematic error of $\delta r |_{\rm
  syst} = 0.01$ 
(i.e. 1\%), the resultant predictions for the m.f.d. per pixel
are plotted as the solid line on the top panels of Figs.3 \& 6. 
We note that, in deriving the expressions above, we
have assumed that $r_j \ll 1$, and so our calculations are not
directly applicable to either Examples A \& B, which have mean flux
ratios per pixel of 0.18 and 0.36. Therefore we expect a closer fit to
the numerical results for Example A than for Example B, which is seen in
Figs.3 \& 6. The predicted m.f.d. per pixel are good estimates
for the calculated values.

\begin{figure}
\vspace{7.8cm}
\includegraphics{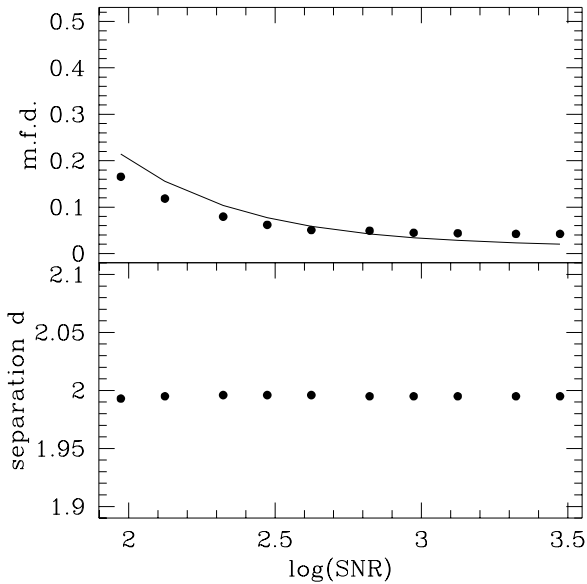}
\caption{Variation of mean fractional deviation per pixel of the
  extracted fluxes from the input fluxes (eq.6) (top panel), and
  best-fit separation (bottom panel) for Example B with increasing
  signal-to-noise ratio (SNR).
  The solid line in the top panel is the m.f.d. per pixel
  expected for this example, with systematic error of 1\% see \S4.
}
\end{figure}

\begin{figure}
\vspace{7.8cm}
\includegraphics{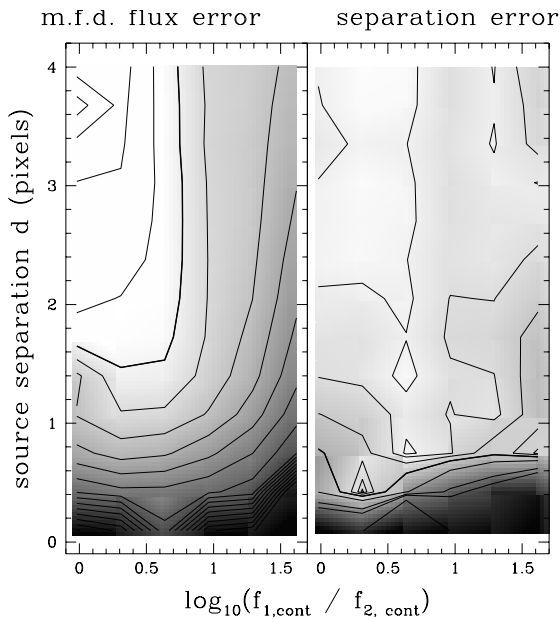}
\caption{Map of mean fractional deviation (left) and error in the best-fit separation (right)
for a set of simulations using the input spectra of Example B, but
  varying the flux ratios and the object separation.
The contours are logarithmically spaced with steps of
0.2dex. The bold contours correspond to an error of 0.1 (i.e. 10\%).
In the left panel, the highest contour corresponds to an 2.6\%
  error and is situated in the top left corner. In the right panel the
  highest contour corresponds to 2.6\% and is in the middle--top left of the
  panel.
}
\end{figure}

%-----------------------------------------------------------------%
\section{Observing practicalities}

The method is clearly able to separate binary object fluxes if the SNR
is high enough. However, there are several practical aspects of
the technique which need to be noted. Whilst some of these are
discussed in detail in Bailey (1998a), and also in Baines (2004) it is
useful to provide a list including aspects gleaned in the development
of the method.

1. Both components of the binary must be {\em fully} in the spectrograph
   slit else unknown fractions of the flux from one component will
   be observed, and so the method will fail.
   Use of a wide slit will degrade the
   spectral resolution of the final observation a little, although
   this is a small sacrifice as it is of paramount importance that
   both objects are fully within the slit. 
   If it is not known {\it a priori} whether the object is a binary or
   not, then a wide slit should be used.

2. Full sampling of the convolved flux distribution is essential -- as
already stated at least 4 pixels (or equivalent for dithering 
patterns) must sample this function to avoid Nyquist sampling problems.

3. For some spectrographs, it is easily possible to repeat the
   observation with the instrument rotated through 180$^\circ$. This
   is useful as it allows any instrumental misalignment problems to be
   subtracted -- this point is particularly stressed in Bailey
   (1998a) and is worth repeating here.

4. The flux separation method will only operate in the spatial
   direction parallel to the slit axis. If the line-of-centres axis is
   orientated normal to the slit axis, then no flux separation will be
   possible. At intermediate angles between these two axes then the
   best-fit separation will be equal to the actual separation
   multiplied by the cosine of the angle between the axes. Information
   on the position angle of the distribution may be obtained by
   combining measurements of two observations taken with the spectrometer
   slit at two different angles on the sky (e.g. see Baines \etal,
   2004). 

5. An important aspect of the method is to observe a single object
   with a similar position on the sky as the intended target in order
   to measure the intrinsic width of the convolution function.
   These single object observations can also be used to calculate series of
   convolutions of 
   binary flux distributions, in order to generate a ``look-up'' table
   of values to be used in the deconvolution, instead of using eq.5.
   In practice, this may be difficult: the seeing can change with
   time, making single star observations less useful. When one object
   has a particularly strong emission line, then the measured width of
   the flux distribution at line centre $\sigma_j$ may be used as an
   estimator for the intrinsic width $\sigma_S$ (see Appendix A.2).

%-----------------------------------------------------------------%
\section{Discussion}

Given the volume of published work dedicated to the deconvolution of
images, why are we investigating new techniques to achieve the same
result?
As the image is sampled in discrete intervals (the pixels), then any
deconvolution method can only increase its resolution (a result from 
sampling theory).
The fundamental aspect of the spectroastrometric technique is that it
utilises changes in the flux distribution convolved with the PSF. As
this convolution takes place {\em before} the distribution is sampled onto
the pixels, then information on scales smaller than the pixels can be
retrieved (this is a different approach to the problem
than deconvolution methods, such as those described in Courbin \etal,
2000).  This is because the information about the binary is spread
throughout the PSF. Therefore, if the PSF itself is well sampled
then extraction of point sources separated by less than the
traditional spatial Nyquist sampling interval (pixel) is possible.
Hence extraction of fluxes is possible at a 
much higher spatial resolution than with previous methods. 
Of course, there is a price to pay for this advantage over
deconvolution techniques. We have assumed {\it a priori} two
point sources, and so have imposed a criterion on the flux
distribution (a feature not present in traditional deconvolution
methods). Also this is essentially an empirical method -- method noise
will always be present in the solution.

Spectroastrometry is likely to achieve best results for the cases when
specific known binary sources are targeted. Serendipitous binary
discovery may prove frustrating when it is clear from the data that a
binary component is 
present, although little else may be derived because, for example, the
secondary component is not fully in the slit, or that the SNR is not
high enough to extract the individual fluxes reliably.
The technique is sufficiently general that any binary object may be
examined (although triple systems may cause problems!).

How powerful is the technique? For ground based telescopes, the
dominant contributor to the point spread function is the seeing, which
typically ranges from one--few arcsec.
With an equivalent signal to
noise ratio of several hundred (often achieved with 
modern instruments)

and pixel sizes of a few tenths of an arcsec (typical
of spectrographs), the position centroid may be determined to
$\sim$milliarcsecond accuracy.
Spectroastrometry should then be a natural complementary
technique to the growing optical/IR interferometry field which
operates at a similar angular scale.

Future investigations into spectroastrometry will concentrate on
continuous flux distributions, such as discs, where the main goal will
be to measure the kinematics of the emitting gas on sub-pixel
scales. This technique is difficult, in that high SNRs are necessary
(and hence the observational aspects may be challenging),
but does promise great rewards.

%-----------------------------------------------------------------%
\section{Conclusion}

We describe the technique of spectroastrometry
and present a method to separate individual binary object fluxes from
the point-spread function which is present in all observations. The
method makes use of the total flux, the position offsets and the
characteristic width of the flux distribution in the spatial
direction: the three observational quantities it is possible to
measure directly from a long slit spectrograph image.
The observing technique to achieve this is not difficult for careful
observers. 

We have demonstrated that the performance of the technique is similar
to previously published deconvolution methods which 
reconstructs the original flux distribution on the sampling
scale ($\sim$pixels). Although spectroastrometry cannot achieve this
complete mapping, it is the only technique in our knowledge, which is
able to successfully separate the fluxes of binary objects with sub-pixel
separations.

The technique of spectroastrometry is a natural counterpart to
interferometry, as it provides spectral information (and hence
kinematics) at similar resolutions, which with available
instrumentation is typically milliarcseconds.

%-----------------------------------------------------------------%
\begin{acknowledgements}
We thanks the anonymous referee for very helpful comments on the submitted
version of this paper.
DB thanks PPARC for postgraduate studentship award.

\end{acknowledgements}

%-----------------------------------------------------------
\appendix

\section{Random and Systematic Errors}
If the width $\sigma_j$ of the spatial distribution of
the flux is sampled over at least two pixels (so the profile will
extend over at least 4 pixels), then the extent of the profile does
not effect the error estimates (essentially this is an expression of
Nyquist sampling). 

The two error sources will be from photon
counting statistics and read-noise.
Where the read-noise errors are dominant in determining the pixel
signal-to-noise ratio (SNR), it can be justified that as long as
the errors are symmetrically distributed around zero 
then their net effect on the
$\mu_j$ and $\sigma_j$ should be zero (especially when robust
location and width estimators are used). 

\subsection{Error in the flux ratio}
When photon noise dominates, the error in the number of photons
$F_{ij}$ in each
pixel is simply $\delta F_{ij} = F_{ij}^{1/2}$.
The signal-to-noise ratio of the position-integrated flux is 
SNR$_j=(\sum_i F_{ij})/ (\sum_i F_{ij})^{1/2} = (\sum_i F_{ij})^{1/2}$.
The standard errors in the position centroid and width (mean and
standard deviation) are the familiar expressions  
$\delta\mu_j^\prime = \sigma_j / n_j^{1/2}$ and $\delta\sigma_j =
\sigma_j / (2n_j)^{1/2}$, where $n_j$ is the number of counts in
the position-integrated profile (e.g. Topping, 1972).  Assuming that
the SNR is the same across the spectrum (i.e. SNR$= $SNR$_j$,
identical for all $j$) we find
\begin{equation}
\delta \mu_j^\prime = \frac{\sigma_j}{{\rm SNR}} \ \ \ {\rm and}\ \ 
\delta \sigma_j = \frac{\sigma_j}{\sqrt{2}{\rm SNR}}.
\end{equation}
The more the seeing function deviates from a Gaussian, then these
expressions for the  
errors $\delta\sigma_j$ and $\delta\mu_j^\prime$ become less accurate 
(e.g. see Beers \etal\ 1990).

For a given error in the position centroid and width we can estimate
the typical error in using the technique.
Let us assume that the random error in the intrinsic width of the seeing
profile is small in comparison with $\delta\sigma_j$, i.e. $\delta
\sigma_S \ll \delta\sigma_j$ (see A.2 for the consequences of
incorrectly deriving $\sigma_S$).
Furthermore, let us assume that the random error in the
separation $\delta d$ is small, and so the dominant error source in
the flux ratio is via the variation in the width $\sigma_j$ (our
method of solution uses the separation $d$ as the independent variable
to minimise the $\chi^2$, and hence $\delta d$ is dominated by
systematic errors).

Rearrangement of eq.5 to make the flux ratio $r_j =
f_{2,j}/f_{1,j}$ the subject allows a simple
error analysis to be performed. This produces the random error 
in the flux ratio $\delta r_j$ of:
\begin{equation}
\left. \delta r_j \right|_{\rm rand} \approx
\frac{1}{a c
  \left(\frac{d}{\sigma_S}\right)^{b}}
\left(\frac{\delta \sigma_j}{\sigma_S}\right),
\end{equation}
where we have also assumed that $r_j \ll 1$, i.e. the case for
small flux ratios.
There will also be a
systematic error in this ratio $\delta r |_{\rm syst}$
related to the inaccuracy in the fitting formula eq.5, leading to the
total error in the flux ratio as
\begin{eqnarray}
\delta r_j &  = & 
\left. \delta r_j \right|_{\rm rand} +
\left. \delta r_j \right|_{\rm syst} \nonumber \\
& \approx &  
\frac{1}{ac}\left(\frac{\sigma_S}{d}\right)^b
\left(\frac{\delta \sigma_j}{\sigma_S}\right) + 
\left. \delta r \right|_{\rm syst} \nonumber \\
 & \approx & 
\frac{1}{\sqrt{2} ac}
\left(\frac{\sigma_S}{d}\right)^b
\left(\frac{1}{{\rm SNR}}\right) + 
\left. \delta r \right|_{\rm syst}
\end{eqnarray}
where, for the final expression, we have inserted eq.A.1.
%The systematic error depends on the value of flux ratio 
%and will lead to an error in the calculation of the
%continuum ratio of the objects.

\subsection{Systematic errors in the intrinsic $\sigma_S$ and its time variation}

We have assumed that the measurement error in the intrinsic
width of the flux distribution $\sigma_S$ is zero.
In practice this may not be the case as there is no
high SNR observation of a single object or where the seeing is
changing over time.
An estimate of $\sigma_S$ could be made from the binary objects'
observations by assuming that the smallest measured value of
$\sigma_j$ occurs when one of the objects completely dominates the
output: this is correct in the limit $f_{2,j}/f_{1,j} \rightarrow 0$, as
then $\sigma_j \rightarrow \sigma_S$.
If this approximation is used, then $\sigma_S$ will
in general be slightly overestimated. What are the consequences from
this assumption?

\begin{figure}
\vspace{7.8cm}
\includegraphics{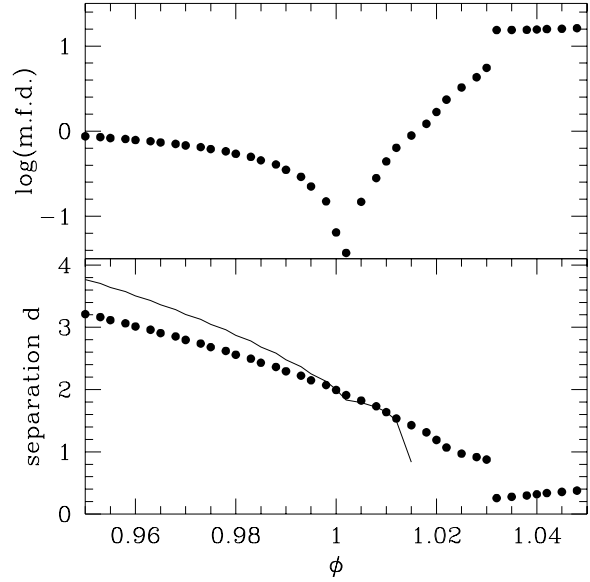}
\caption{Top panel: logarithm of the m.f.d. per pixel versus
  the systematic error in the intrinsic width of the seeing
  $\sigma_S$. Bottom panel: the variation of the best fit
  separation. See text.
}
\end{figure}

Both the best-fit separation between the components, and the
individual fluxes (notably the continuum flux ratio) will have an
associated error.
From examination of eq.5 we see that the fractional change in the
separation caused solely by an change in $\sigma_S$ ($\delta
\sigma_S$) is
\begin{equation}
\frac{\delta d}{d} = \frac{\delta \sigma_S}{\sigma_S} 
\left[ -\frac{1}{b} \left( \frac{\sigma_S}{\sigma_j - \sigma_S}\right) 
+ 1 - \frac{1}{b} \right]
\end{equation}
which assumes that there is no resultant error in the flux ratio
(strictly this is never the case, but we proceed in the spirit of
producing relatively simple error estimates).
Note that an overestimate of $\sigma_S$ leads to a systematic
underestimate of $d$.

We have calculated a series of simulations for Example A in the text
with high SNR of 3000, and systematically changed the value of
$\sigma_S$ used in the flux separation. We take the actual value of
$\sigma_S$ and multiply by a fraction $\phi$ ranging from 0.95--1.05
to mimic both an underestimate and an overestimate of $\sigma_S$ from
the data. 
The results are shown as the filled circles in Fig.A1: the top panel
is the logarithm of the m.f.d. per pixel, and the bottom panel is
the best-fit separation between the components. It is clear that $\phi
\approx 1$ (close to unity) produces the best results by far. This
underlines the desirability for accurate measurement of $\sigma_s$.
When $\phi > 1.03$, the numerical method fails, as all the 
values of $(\sigma - \sigma_S)/\sigma_S < 0.0$. This produces a
discontinuity in the trends in Fig.A.1.

Plotted as a solid line on the bottom panel is 
our estimate (above, with $b = 1.93$) for the variation in $d$.
For each value of $\phi$ we calculate a mean value of $(\sigma -
\sigma_S)/\sigma$ per pixel for eq.A.4 (this produces the main
difference between our estimated values from eq.A.4.and the numerical
values).
The calculated value predicts  a zero
crossing in $d$ at $\phi = 1.017$, as there the {\em mean} value of
$(\sigma-\sigma_S)/\sigma_S$ is zero.

\end{document}